\documentclass[runningheads]{llncs}
\usepackage{graphicx}
\begin{document}

\title{Query expansion with artificially generated texts}
%
%\titlerunning{Abbreviated paper title}
% If the paper title is too long for the running head, you can set
% an abbreviated paper title here
%
\author{Vincent Claveau\orcidID{0000-0002-3459-0550}}
\authorrunning{Vincent Claveau}
% First names are abbreviated in the running head.
% If there are more than two authors, 'et al.' is used.
%
\institute{CNRS - IRISA, Univ. Rennes\\Campus de Beaulieu, F-35042 Rennes, France\\
\email{vincent.claveau@irisa.fr}}
\maketitle              % typeset the header of the contribution
\begin{abstract}
A well-known way to improve the performance of document retrieval is to expand the user's query. Several approaches have been proposed in the literature, and some of them are considered as yielding state-of-the-art results in IR. 
%In the traditional IR setting, a user expresses his information needs with the help of a query. Yet, matching these usually short queries with documents i It is known that expanding or enriching the query often helps to yield better results, and several approaches have been proposed.
In this paper, we explore the use of text generation to automatically expand the queries. 
We rely on a well-known neural generative model, GPT-2, that comes with pre-trained models for English but can also be fine-tuned on specific corpora.
Through different experiments, we show that text generation is a very effective way to improve the performance of an IR system, with a large margin (+10\% MAP gains), and that it outperforms strong baselines also relying on query expansion (LM+RM3). 
This conceptually simple approach can easily be implemented on any IR system thanks to the availability of GPT code and models. 

\end{abstract}

%%%%%%%%%%%%%%%%%%%%%%%%%%%%%%%%%%%%%%%%%%%%%%%%%%%%%%%%%%%%%%%%%%%%%%%%%%%%%%%%%%%%%%%%%%
%%%%%%%%%%%%%%%%%%%%%%%%%%%%%%%%%%%%%%%%%%%%%%%%%%%%%%%%%%%%%%%%%%%%%%%%%%%%%%%%%%%%%%%%%%

\section{Introduction}

In the IR traditional setting, a user expresses his information needs with the help of a query. Yet, it is sometimes difficult to match the query with the documents, for instance because of the query vocabulary may differ from the documents. Especially when the query is short, the performance of the system is usually poor, as it is difficult to detect the precise focus of the information need, and the relative importance of the query terms. 

Query expansion aims at tackling these problems by transforming the short query into a larger text (or set of words) that make it easier to match documents from the collection. 
The main difficulty of query expansion is obviously to add only relevant terms to the initial query. Several techniques have been proposed in the literature, based on linguistic resources (e.g. synonym lists) or based on the documents themselves (e.g. pseudo-relevance feedback).

In this paper, we explore the use of recent text generation models to expand queries. We experimentally demonstrate that the recent advances in neural generation can dramatically improve ad-hoc retrieval, even when dealing with specialized domains. More precisely, through different experiments, we show that:
\begin{enumerate}
\item texts artificially generated from the query can be used for query expansion;
\item this approach does not only provide new terms to the query, but also a better estimate of their relative weights;
\item in addition, it also provides a better estimate of the importance (i.e. weight) of original query words;
\item this approach can also be used on specialized domains.
\end{enumerate}

The paper is structured as follows. After a presentation of the related work, Section~\ref{sec:approach} details the different components of our approach. Several experiments are then detailed in Section~\ref{sec:expes}. Last, some concluding remarks are given in Section~\ref{sec:concl}.

%%%%%%%%%%%%%%%%%%%%%%%%%%%%%%%%%%%%%%%%%%%%%%%%%%%%%%%%%%%%%%%%%%%%%%%%%%%%%%%%%%%%%%%%%%

\section{Related work}

Query expansion is a well-established technique to try to improve the performance of an IR system. Adding new terms to the query is expected to specifically improve recall, yet, since the query is, hopefully, better formulated, it may also improve the top rank results and be beneficial to precision.
One might classify the existing automatic approaches based on the origins of resource used to expand the query.

\paragraph{External resources.}

One obvious way to expand a query is to add semantically related terms to it (synonyms or sharing other semantic relations like hyponyms, quasi-synonyms, meronyms...). Existing lexical resources can be used to add, for each query term, a list of semantically related terms; yet, one has to deal with different problems: existence of lexical resources for the collection language, or for the specific domain of the collection, choice of the appropriateness of certain relations, need of sense disambiguation for polysemous words...
WordNet~\cite{miller-ijl-90} is among the best-known resources for English (general language) and have been used with mitigated results at first \cite{Voorhees94}, but later shown to be effective \cite[inter alia]{Claveau-COLING2016}.

\paragraph{Collection-based resources.}
Distributional thesauri have also been exploited to enrich queries. Since they can be built from the document collection (or from a large corpus with similar characteristics), they are suited to the domain, the vocabulary...
Traditional techniques to build these thesauri have obtained good results for query expansion \cite{Claveau-COLING2016}.
Neural approaches, that is, word embedding approaches are now widely used to build such semantic resources. 
In the recent years, static embeddings (word2vec \cite{Mikolov13}, Glove \cite{pennington2014glove} or FastText \cite{bojanowski2016enriching} to name a few) were also used in IR, in particular to enrich the query. Indeed, these trainable dense representations make it easy to find new words that are semantically close to query words.

Even more recently, dynamic word representations obtained with transformer-based architectures, such as BERT \cite{devlin2019bert} or GPT \cite{Radford2019}, have been proposed. They build a representation for each word according to  its context, and this ability have been exploited to obtain competitive results in IR tasks \cite[inter alia]{Dai_2019,Khattab2020}.
Bert has been also used for query expansion in the framework of a neural IR system based on reranking \cite{Padaki2020}

\paragraph{Pseudo-relevance feedback}
A last category of studies considers only a small set of document to help to expand or reformulate the query. To be automatic, they replace the user feedback by the hypothesis that the best ranked documents retrieved with the original query are relevant and may contain useful semantic information \cite{Ruthven03}. It is interesting to note that in this case, not only semantically relevant terms are extracted, but also distributional/statistical information on them and on the original query terms.
In this category, Rocchio, developed in the 60's for vector space model was among the first one popularized \cite{Manning_IR2008}. One of the current best known approach is RM3, which was developed in the framework of language model based IR systems \cite{Abdul-RM3}. It is often reported to yield the best results in ad-hoc retrieval tasks, even compared with recent neural models \cite{NeuralHype}. 
Neural approaches have also been proposed to integrate pseudo-relevance feedback information \cite{Li-NPRF-EMNLP2018}, yet, as it is reported by the authors, the results are still lower than traditional models with query expansion.

~\newline

In this paper, we propose to use constrained text generation to expand queries. In this approach, the original query is used as a seed for a generative model which will output texts that are, hopefully, related to the query. 
While text generation with language model is not new, the performance of neural models based on transformers \cite{Vaswani_NIPS2017} makes this task realistic. In this paper, we use the Generative Pre-Trained Transformers (GPT). They are built from stacked transformers (precisely, decoders) trained on a large corpus by auto-regression, i.e. unsupervisedly trained to predict the next token given the previous ones. 
The second version, GPT-2 \cite{Radford2019}, contains 1.5 billion parameters for its largest pre-trained model, trained on more than 8M documents from Reddit (i.e. mostly English and general language such as discussion on press articles).
A newer version, GPT-3, has been released in July 2020; it is much more larger (175 billion parameters) and outperforms GPT-2 on any tested task. Yet, the experiments reported below were done before this release, thus with GPT-2.

%%%%%%%%%%%%%%%%%%%%%%%%%%%%%%%%%%%%%%%%%%%%%%%%%%%%%%%%%%%%%%%%%%%%%%%%%%%%%%%%%%%%%%%%%%

\section{Generated query expansion}
\label{sec:approach}

\subsection{Overview of our approach}

As it was previously explained, our approach is very simple as it relies on existing tools and techniques. 
From a query, multiple texts are generated by a GPT-2 model using the query as the seed. Note that the generation process is not deterministic, and thus, even with the same seed, the texts are different. In the experiments reported below, 100 texts per query are generated (unless specified otherwise).
These texts are concatenated and considered as the new query. 
In our experiments, this new, very large, query is then fed to a simple BM25+ IR system, but it could obviously be used in any other IR system.

An example of a text generated from a query is presented in Fig.~\ref{fig:example_701}. As one can see, the generated text, while completely invented (note the barrel prices), is relevant for the query. It contains many terms, absent from the original query, that are more or less closely related to the information need, such as orthographic variants (United States vs. U.S.), meronyms-metonyms (barrel vs. oil), hypernyms (energy vs. oil) and more generally any paradigmatic (consumer, producer vs. industry) or syntagmatic (production for oil) relations.
It is worth noting that such texts also give a valuable information about the relative frequency of each terms (contrary to thesauri or embeddings).

\begin{figure}
    \noindent\fbox{%
    \parbox{\textwidth}{%
U.S. oil production has been declining steadily for decades and it is not expected to reverse. In fact, some argue that it may even get worse. The long-term trend is for oil production to decline at a rate of about 1 percent per year. With production of about 8 million barrels per day now, there is no way the United States can replace its current output.

The U.S. oil boom was a result of an energy revolution in the 1970s that led to increased oil production, and a significant change in the global oil market. The U.S. now produces about 2.3 million barrels of oil per day, the highest it has been in over 30 years.

The United States is now the world's largest oil producer and the fourth largest oil exporter.

What happened?

When oil prices peaked in the 1970s, the United States was the world's largest oil producer. But over the next several decades, the United States' oil production began to decline. The decline was most pronounced in the 1980s, when the United States began to fall behind other oil producing countries.

The oil price decline in the 1970s was not entirely voluntary. The United States was producing less oil and exporting more oil than it was consuming. The Federal Reserve controlled the amount of dollars in the Federal Reserve's reserves, so the United States was not exporting as much oil as it was producing. The decline in U.S. oil production was a result of the declining price of oil.

The price of oil had declined from \$8 per barrel in 1973 to \$2.50 per barrel in 1977. In 1979, the price of oil reached a high of \$15.75 per barrel. By 1983, the price of oil had fallen to \$4.65 per barrel. By 1986, the price of oil had fallen to \$1.86 per barrel. By the end of the 1980s, the price of oil had fallen to \$1.24 per barrel.

The decline in oil prices was a direct result of the energy revolution in the 1970s. The United States was the world's largest oil producer, but the United States was also the world's largest consumer of oil. When oil prices fell, so did the cost of producing oil. %
}}
    \caption{Example of a document generated with the pre-trained GPT-2 large model from the text seed "U.S. oil industry history" (query 701 from .GOV collection)}
    \label{fig:example_701}
\end{figure}

\subsection{Pre-trained models, fine-tuning and parameters}
\label{sec:fine_tuning}

GPT-2 comes with several pre-trained models, having different size in terms of parameters (from 124M to 1.5B). As it was previously said, their training data was news-oriented general language. The largest model was used for two of the tested collections (see below).
While these all-purpose models are fine for IR collections whose documents are also general language, it may not be appropriate for domain-specific IR collections. In the experiment reported in the next section, we use the \textsc{ohsumed} collection, made of medical documents.  
For this collection, we have fine-tuned the GPT-2 355M model on the documents of the collection in order to adapt the language model to the specific medical syntax and vocabulary. The fine-tuning was stopped after 250,000 samples were processed (this number of sample process indirectly controls the under/over-fitting to the specialized corpus) was set to  and other parameters (batch size, optimizer, learning rate...) let to their defaults. Although a larger set of medical documents could be used (from Pubmed for instance), this small fine-tuned model is expected to be more suited to generate useful documents to enrich the query.

Concerning the generation of documents, for reproducibity purposes, here are the main GPT-2 parameters used (please refer to GPT-2 documentation\footnote{\url{https://github.com/openai/gpt-2}}): 
length=512, temperature=0.5, top\_p=0.95, top\_k = 40.

\subsection{IR Systems}

In the experiments reported in the next section, we use two IR models. 
The first one is BM25+ \cite{Lv2011}, a variant  of BM25 \cite{RWB98}. The parameters $k_1$, $k_3$, $b$ and $\delta$  were kept at their default value (resp. 1.2, 1000, 0.75, 1). It is implemented as a custom modification of the \textsc{gensim} toolkit \cite{Rehurek_GENSIM}.
The second IR model is Language modeling with Dirichlet smoothing \cite{Zhai2001} as implemented in Indri \cite{MetzlerCroft2004,StrohmanMetzlerTurtleCroft2005}. The smoothing parameter $\mu$ is set to 2\,500.
Both models are regarded as yielding state-of-the-art performance for bag-of-words representation \cite{NeuralHype}.
Their RSV function can be written:
$$RSV(q,d) =\sum_{t \in q} w_q(t) \cdot w_d(t) $$
with  $w_q(t)$ the weight of term $t$ in query $q$ and $w_d(t)$ the weight in document $d$, as illustrated in Tab.~\ref{tab:weight} (from \cite{Lv2011}).
\begin{table}
\begin{center}
\begin{tabular}{l|cc}
IR model & weighting\\
 \hline
BM25+  $w_d(t)$  & $\left( \frac{(k1+1)c(t,d)}{k1(1-b+b\cdot dl(d)/avdl)+c(t,d)}+\delta \right) \cdot \log \frac{N+1}{df(t)}$\\
BM25+ $w_q(t)$  & $\frac{(k3+1)c(t,q)}{k3+c(t,q)}$ \\
        &  with $k1, k_3, b$ and $\delta$ fixed parameters \\
       & \\

LM $w_d(t)$   & $\log \left( \frac{\mu}{dl(d)+\mu} + \frac{c(t,d)}{(dl(d)+\mu)p(t|C)} \right)$\\
LM $w_q(t)$   & $c(t,q)$ \\
              & $\mu>0$ a smoothing parameter\\
       &\\
 
\end{tabular}
\end{center}
\caption{\label{tab:weight}IR models (weighting functions of terms in the query and the document) for BM25+ \protect\cite{RWB98,Lv2011} and Language modeling with  Dirichlet smoothing LM \protect\cite{Zhai2001}}
\end{table}
For RM3 expansion, we also rely on the Indri implementation; the results reported in the next section corresponds to the best performing parameters tested for each collection (number of documents considered for pseudo relevance feedback and number of terms kept).

%%%%%%%%%%%%%%%%%%%%%%%%%%%%%%%%%%%%%%%%%%%%%%%%%%%%%%%%%%%%%%%%%%%%%%%%%%%%%%%%%%%%%%%%%%

\section{Experiments}
\label{sec:expes}

\subsection{Experimental settings}

Three IR collections are used in our experiments: Tipster, GOV2 and  \textsc{ohsumed}. Some basic statistics are given in Tab.~\ref{tab:collection}.

\begin{table}[tb]
\begin{center}
%\scalebox{0.9}{
\begin{tabular}{l|ccc}
                     & Tipster & GOV2 &\textsc{ohsumed} \\ 
\hline
nb of documents      & 170,000 & 25M  & 350,000 \\ 
nb of queries        & 50      & 150  & 106 \\
avg size of queries  & 6.74    & 3.15 & 7.24\\
language             & En      & En   & En \\
avg nb of relevant doc per query & 849  & 179 & 21 \\
\end{tabular} %}
\end{center}
\caption{\label{tab:collection}Statistics on the IR collections used}
\end{table}

Tipster was used in TREC-2. The documents are articles from newspaper, patents and specialized press (computer related) in English. The queries are composed of several parts, including the query itself and a narrative detailing the relevance criteria; in the experiments reported below, only the actual query part is used.

GOV2 is a large collection of  Web pages crawled from the .gov domain and used in several TREC tracks. In the experiments reported below, 150 queries from TREC 2004-2006 ad-hoc retrieval tasks are used; as for Tipster, only the actual query part is used (i.e. description and narrative fields are not included in the query).

\textsc{Ohsumed} \cite{Ohsumed94} contains bibliographical notices from Medline and queries from the TREC-9 filtering task. Its interest for our experiments is that it deals with a specialized domain, hence it contains a specific vocabulary.

Performance are assessed with standard scores: Precision at different thresholds (P@x), R-precision (R-prec), \textit{Mean Average Precision} (MAP). When needed, a paired t-test with $p=0.05$ is performed to assess the statistical significance of the difference between systems.

\subsection{General language}

Tables~\ref{tab:res-tipster} and \ref{tab:res-gov2} respectively present the results for the general-language collections Tipster and GOV2.
For comparison purposes, we indicate the results of BM25+ without expansion, Indri's Language Model (LM) with and without RM3 expansion. The best performing setting for RM3 on Tipster is 100 terms from the top 20 documents, and 100 terms for the top 10 documents GOV2. The statistical significance is computed by comparing with the LM+RM3 baseline.

\begin{table}[tb]
\begin{center}
%\scalebox{0.9}{
\begin{tabular}{l|ccccccc}
                        & MAP    &  R-Prec &  P@5   & P@10  & P@20 & P@100 \\
\hline
BM25+                   & 25.06  &  32.16 & 95.60 & 92.60 & 89.70 & 73.64 \\
LM                      &  24.48 &  31.48 & 92.40 & 89.00 & 85.40 & 70.70 \\
LM + RM3                & 31.01  & 36.38  & 94.40 & 93.20 & 90.60 & 81.22 \\
\hline

BM25+ and expansion      &  \textbf{35.22}$^*$ & \textbf{39.87}$^*$   & \textbf{99.60}$^*$ & \textbf{98.40}$^*$ & \textbf{98.20}$^*$ & \textbf{87.84}$^*$ \\
\end{tabular} %}
\end{center}
\caption{\label{tab:res-tipster}Performance (\%) on Tipster with query expansion; best results in bold, statistical significance over LM+RM3 noted with $*$}
\end{table}

\begin{table}[tb]
\begin{center}
%\scalebox{0.9}{
\begin{tabular}{l|ccccccc}
                        & MAP    &  R-Prec & P@5   & P@10  & P@20 & P@100 \\
\hline
BM25+                   & 25.66  &  31.25  & 52.92 & 49.97 & 46.52 & 34.63 \\
LM                      & 27.96  &  33.01  & 56.08 & 55.20 & 51.59 & 37.32 \\
%LM with RM3             & 29.16  &  32.87  & 54.86 & 54.53 & 50.34 & 37.65 \\ % k = 10; nb terms = 10 
LM with RM3             & 30.22  &  34.20  & 55.00 & 56.08 & 53.67 & \textbf{45.86} \\ % k = 10; nb terms = 100 
\hline

BM25+ and expansion     &  \textbf{34.54}$^*$ &  \textbf{37.76}  & \textbf{67.91}$^*$ & \textbf{63.88}$^*$ & \textbf{57.94} & 44.30 \\

\end{tabular} % }
\end{center}
\caption{\label{tab:res-gov2}Performance (\%) on GOV2 with query expansion; best results in bold, statistical significance over LM+RM3 noted with $*$}
\end{table}

On both collections, and on every performance measure, expanding the queries with the generated texts brings important gains compared with the system without expansion. Also, our approach outperforms RM3 expansion in almost every situation, and with a large margin on MAP, R-prec and precision on the top-ranked documents (P@5, P@10).

\subsection{Specialized language}

The same setting is used on the \textsc{ohsumed} collection. For these medical-oriented IR dataset, we report two versions of our approach: one is using the pre-trained model as before, and one relies on a model fine-tuned on the documents of the collection. The best performing setting for RM3 is 80 terms for the top 10 documents. The results are reported in Tab.~\ref{tab:res-ohsumed}. 

\begin{table}[tb]
\begin{center}
%\scalebox{0.9}{
\begin{tabular}{l|ccccccc}
                                     & MAP    &  R-Prec & P@5   & P@10  & P@20 & P@100 \\
\hline
BM25+                                & 18.27  &   19.94 & 31.88 & 26.04 & 20.50 & 9.48 \\
LM                                   & 17.61 &    20.35 & 29.31 & 24.06 & 19.21 & 9.28 \\
LM + RM3                             & 20.80 &    22.54 & 30.89 & 26.83 & 22.18 & 10.51 \\
\hline
BM25+ and expansion (no fine-tuning) & 21.60 & 23.75 & 33.47$^*$ & 27.62 & 22.92 & 11.16\\
BM25+ and expansion (fine-tuning)    & \textbf{23.07}$^*$ & \textbf{24.65}$^*$ & \textbf{34.65}$^*$ & \textbf{29.41}$^*$ & \textbf{24.31} & \textbf{11.42} \\

\end{tabular} %}
\end{center}
\caption{\label{tab:res-ohsumed}Performance (\%) on \textsc{ohsumed} with query expansion; best results in bold, statistical significance over LM+RM3 noted with $*$}
\end{table}

Here again, the GPT-based expansion significantly improves the results of the IR system and outperforms RM3 expansion.
Yet, the gains are lower than for the two previous collections. This difference can be explained by the following factors:
\begin{enumerate}
    \item the queries are longer more complex and more specific (as can be seen in Tab.~\ref{tab:collection}, few documents are relevant); 
    \item the generation model is not sufficiently suited to the documents. 
\end{enumerate}
Concerning this latter reason, we can indeed see the interest of fine-tuning the generation model, but better results may be obtained by using a larger set of medical documents, or adopting different fine-tuning parameters (in particular the number of epoch/samples processed, see Sect.~\ref{sec:fine_tuning}). Unfortunately, defining a priori the best parameters for our IR task is not possible and the cost of the fine-tuning process makes it impossible to test a wide range of possible values.

\subsection{Query expansion and weight information}

\begin{figure}
    \centering
    \includegraphics[width=0.75\textwidth]{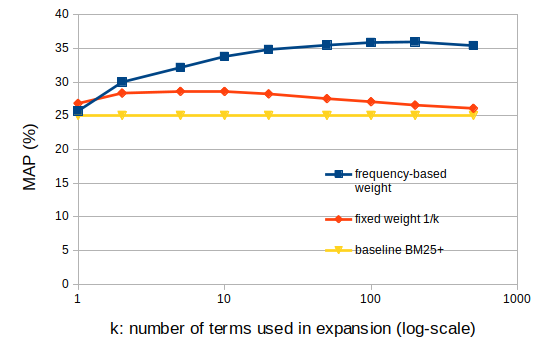}
    \caption{MAP (\%) according to size of the expansion, terms with fixed weight or weight depending on their frequency in the generated documents; Tipster collection}
    \label{fig:exp_size}
\end{figure}

One of the interest of having complete texts that are generated is that we can collect information on the relative importance of words, to the contrary of expanding queries with a thesaurus. 
To observe the impact of the number of occurrences in the generated texts, we evaluate the effect of keeping the $k$ most frequent terms of the generated texts and either weighting them by their frequency (as done usually by BM25) or by giving a fixed weight ($1/k$). The results for different values of $k$ are presented in Fig.~\ref{fig:exp_size}.
On can observe that adding terms to the query with a fixed weight slightly improves the MAP, but most of the gain is indeed brought by a proper weighting based on the frequency of the term in the generated documents.

In the next experiment, we examine how the generated texts can help to re-weight the initial query terms; there is no query expansion, since only the initial query terms are kept, but their frequency in the generated texts are used in the BM25+ $w_q$. The results reported in Tab.~\ref{tab:res-reweighting} show that there is indeed a small improvement of the MAP, that is more noticeable at high DCV.
\begin{table}[tb]
\begin{center}
%\scalebox{0.9}{
\begin{tabular}{l|ccccccc}
                        & MAP    &  R-Prec &  P@5   & P@10  & P@20 & P@100 \\
\hline
BM25+                   & 25.06  &  32.16 & 95.60 & 92.60 & 89.70 & 73.64 \\
%BM25+ and expansion     &  35.22 & 39.87  & 99.60 & 98.40 & 98.20 & 87.84 \\
\hline
BM25+ with re-weighting  & 27.12  & 33.22 &  96.80 & 94.20 & 92.70 & 77.12 \\

\end{tabular} %}
\end{center}
\caption{\label{tab:res-reweighting}Performance (\%) on Tipster with re-weighting query words}
\end{table}
These two experiments demonstrate the usefulness of dealing with full texts and not only word-to-word similarity.

\subsection{Number of generated documents}

\begin{figure}
    \centering
    \includegraphics[width=0.7\textwidth]{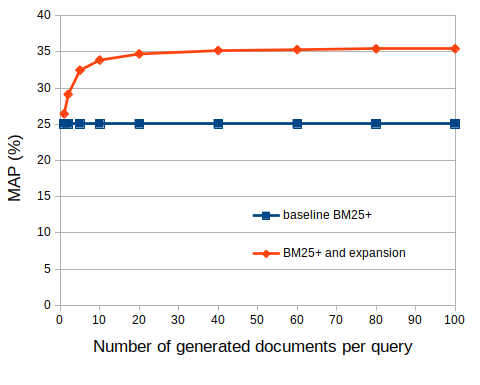}
    \caption{MAP (\%) according to the number of generated documents (average over 5 runs); Tipster collection}
    \label{fig:exp_nb_doc}
\end{figure}

Since text generation can be costly, it is interesting to see how many generated texts are necessary. In Fig.~\ref{fig:exp_nb_doc}, the MAP obtained for up to 100 documents is presented. One can observe that a plateau is rapidly reached (at around 20 documents per query). Of course, the size of the generated documents (can be set as a parameter of the generation process) is also to be considered.

%%%%%%%%%%%%%%%%%%%%%%%%%%%%%%%%%%%%%%%%%%%%%%%%%%%%%%%%%%%%%%%%%%%%%%%%%%%%%%%%%%%%%%%%%%

\section{Conclusive remarks and foreseen work}
\label{sec:concl}

Neural approaches are increasingly used in IR, with mitigated results, especially when compared with "traditional" bag-of-word approaches \cite{NeuralHype,NeuralHype2}. Here, the neural part is successfully used outside of a "traditional" IR system (but note that it could be used with any IR systems, since it simply enriches the query).
The expansion approach presented in this paper is simple and easy to implement (thanks to the availability of the GPT models and code) while offering impressive gains. 
Lot of parameters could be further optimized, especially on the GPT model side (to influence the "creativity" of the text generation), and the fine-tuning capabilities should also be explored more thoroughly (influence of bigger specialized corpus if available, precise mix between pre-trained and fine-tuning, etc.). 
The recent availability of GPT-3\footnote{\url{https://github.com/openai/gpt-3}} makes it possible to even get greater gains thanks to the alleged high quality of its outputs.

This whole approach also offers many research avenues: in this work, we have used text generation as a way to perform data augmentation on the query side, but it could also be used to augment the representation of the documents (even if in practice, the cost is still prohibitive on large collection).
All machine-learning (neural or not) approaches based on pseudo-relevance feedback to train their model could instead use similar text generation with the advantage that they would not be limited by the number of potential relevant documents in the shortlist.
And of course, similar data-augmentation strategy could be used for other tasks than document retrieval.

% additivity of expansion (ie perform PRF on these results)

% generation time

More fundamentally, the recent improvements of text generation also question the relevance of the document retrieval task. Indeed, it is possible to envision systems that will be able to generate one unique document answering the user's information need, similarly to question-answering. If the generative model is trained on the document collection, the generated document will serve as a summary (which is one of the popular applications of GPT-x models) of the relevant documents. 
Yet, the current limitations of the models tested in this paper make them far from being suited for this ultimate task: the generated documents do deal with the subject of the query, and thus use a relevant vocabulary, but do not provide accurate, factual information (as seen in the Example in Fig.~\ref{fig:example_701} about the price of oil barrels).

%\subsubsection*{Acknowledgments.} 
%
%This work was partly funded by a French government support granted to the CominLabs LabEx managed by the ANR in \textit{Investing for the Future} program under reference ANR-10-LABX-07-01.
%  Nous tenons à remercier Laurent Amsaleg (IRISA-CNRS) et Teddy Furon (Inria Rennes) pour l'idée à l'origine de ce travail et les discussions fructueuses tenues avec eux sur l'adaptation de leur approche aux métriques de RI.

% include your own bib file like this:
\bibliographystyle{splncs04}
\bibliography{biblio_gpt4ir.bib}

\begin{thebibliography}{10}
\providecommand{\url}[1]{\texttt{#1}}
\providecommand{\urlprefix}{URL }
\providecommand{\doi}[1]{https://doi.org/#1}

\bibitem{Abdul-RM3}
Abdul-jaleel, N., Allan, J., Croft, W.B., Diaz, O., Larkey, L., Li, X.,
  Smucker, M.D., Wade, C.: Umass at trec 2004: Novelty and hard. In: In
  Proceedings of TREC-13 (2004)

\bibitem{bojanowski2016enriching}
Bojanowski, P., Grave, E., Joulin, A., Mikolov, T.: Enriching word vectors with
  subword information. arXiv preprint arXiv:1607.04606  (2016)

\bibitem{Claveau-COLING2016}
Claveau, V., Kijak, E.: Direct vs. indirect evaluation of distributional
  thesauri. In: Proceedings of {COLING} 2016, the 26th International Conference
  on Computational Linguistics: Technical Papers. pp. 1837--1848. The COLING
  2016 Organizing Committee, Osaka, Japan (Dec 2016),
  \url{https://www.aclweb.org/anthology/C16-1173}

\bibitem{Dai_2019}
Dai, Z., Callan, J.: Deeper text understanding for ir with contextual neural
  language modeling. Proceedings of the 42nd International ACM SIGIR Conference
  on Research and Development in Information Retrieval  (Jul 2019).
  \doi{10.1145/3331184.3331303},
  \url{http://dx.doi.org/10.1145/3331184.3331303}

\bibitem{devlin2019bert}
Devlin, J., Chang, M.W., Lee, K., Toutanova, K.: Bert: Pre-training of deep
  bidirectional transformers for language understanding (2019)

\bibitem{Ohsumed94}
Hersh, W., Buckley, C., Leone, T.J., Hickam, D.: Ohsumed: An interactive
  retrieval evaluation and new large test collection for research. In:
  Proceedings of the 17th Annual International ACM SIGIR Conference on Research
  and Development in Information Retrieval. pp. 192--201. SIGIR '94,
  Springer-Verlag New York, Inc., New York, NY, USA (1994),
  \url{http://dl.acm.org/citation.cfm?id=188490.188557}

\bibitem{Khattab2020}
Khattab, O., Zaharia, M.: Colbert: Efficient and effective passage search via
  contextualized late interaction over {BERT}. In: Huang, J., Chang, Y., Cheng,
  X., Kamps, J., Murdock, V., Wen, J., Liu, Y. (eds.) Proceedings of the 43rd
  International {ACM} {SIGIR} conference on research and development in
  Information Retrieval, {SIGIR} 2020, Virtual Event, China, July 25-30, 2020.
  pp. 39--48. {ACM} (2020). \doi{10.1145/3397271.3401075},
  \url{https://doi.org/10.1145/3397271.3401075}

\bibitem{Li-NPRF-EMNLP2018}
Li, C., Sun, Y., He, B., Wang, L., Hui, K., Yates, A., Sun, L., Xu, J.: {NPRF}:
  A neural pseudo relevance feedback framework for ad-hoc information
  retrieval. In: Proceedings of the 2018 Conference on Empirical Methods in
  Natural Language Processing. pp. 4482--4491. Association for Computational
  Linguistics, Brussels, Belgium (Oct-Nov 2018). \doi{10.18653/v1/D18-1478},
  \url{https://www.aclweb.org/anthology/D18-1478}

\bibitem{NeuralHype}
Lin, J.: The neural hype and comparisons against weak baselines. {SIGIR} Forum
  \textbf{52}(2),  40--51 (2018). \doi{10.1145/3308774.3308781},
  \url{https://doi.org/10.1145/3308774.3308781}

\bibitem{Lv2011}
Lv, Y., Zhai, C.: Lower-bounding term frequency normalization. In: Proc. of the
  20th ACM International Conference on Information and Knowledge Management.
  pp. 7--16. CIKM '11, ACM, New York, NY, USA (2011).
  \doi{10.1145/2063576.2063584},
  \url{http://doi.acm.org/10.1145/2063576.2063584}

\bibitem{Manning_IR2008}
Manning, C.D., Raghavan, P., Sch\"{u}tze, H.: Introduction to Information
  Retrieval. Cambridge University Press, USA (2008)

\bibitem{MetzlerCroft2004}
Metzler, D., Croft, W.: Combining the language model and inference network
  approaches to retrieval. Information Processing and Management Special Issue
  on Bayesian Networks and Information Retrieval  \textbf{40}(5),  735--750
  (2004)

\bibitem{Mikolov13}
Mikolov, T., Sutskever, I., Chen, K., Corrado, G.S., Dean, J.: Distributed
  representations of words and phrases and their compositionality. In: Burges,
  C.J.C., Bottou, L., Ghahramani, Z., Weinberger, K.Q. (eds.) Advances in
  Neural Information Processing Systems 26: 27th Annual Conference on Neural
  Information Processing Systems 2013. Proceedings of a meeting held December
  5-8, 2013, Lake Tahoe, Nevada, United States. pp. 3111--3119 (2013),
  \url{http://papers.nips.cc/paper/5021-distributed-representations-of-words-and-phrases-and-their-compositionality}

\bibitem{miller-ijl-90}
Miller, G.A.: {Word{N}et: An On-Line Lexical Database}. International Journal
  of Lexicography  \textbf{3}(4) (1990)

\bibitem{Padaki2020}
Padaki, R., Dai, Z., Callan, J.: Rethinking query expansion for bert reranking.
  In: Jose, J.M., Yilmaz, E., Magalh{\~a}es, J., Castells, P., Ferro, N.,
  Silva, M.J., Martins, F. (eds.) Advances in Information Retrieval. pp.
  297--304. Springer International Publishing, Cham (2020)

\bibitem{pennington2014glove}
Pennington, J., Socher, R., Manning, C.D.: Glove: Global vectors for word
  representation. In: Empirical Methods in Natural Language Processing (EMNLP).
  pp. 1532--1543 (2014), \url{http://www.aclweb.org/anthology/D14-1162}

\bibitem{Radford2019}
Radford, A., Wu, J., Child, R., Luan, D., Amodei, D., Sutskever, I.: Language
  models are unsupervised multitask learners. OpenAI Blog  (2019)

\bibitem{Rehurek_GENSIM}
{\v R}eh{\r u}{\v r}ek, R., Sojka, P.: {Software Framework for Topic Modelling
  with Large Corpora}. In: {Proceedings of the LREC 2010 Workshop on New
  Challenges for NLP Frameworks}. pp. 45--50. ELRA, Valletta, Malta (May 2010),
  \url{http://is.muni.cz/publication/884893/en}

\bibitem{RWB98}
Robertson, S.E., Walker, S., Hancock-Beaulieu, M.: Okapi at {TREC-7}:
  {A}utomatic {A}d {H}oc, {F}iltering, {VLC} and {I}nteractive. In: Proc. of
  the 7\textsuperscript{th} Text Retrieval Conference, TREC-7. pp. 199--210
  (1998)

\bibitem{Ruthven03}
Ruthven, I., Lalmas, M.: A survey on the use of relevance feedback for
  information access systems. Knowledge Eng. Review  \textbf{18}(2),  95--145
  (2003)

\bibitem{StrohmanMetzlerTurtleCroft2005}
Strohman, T., Metzler, D., Turtle, H., Croft, W.: Indri: A language-model based
  search engine for complex queries (extended version). Tech. rep., CIIR (2005)

\bibitem{Vaswani_NIPS2017}
Vaswani, A., Shazeer, N., Parmar, N., Uszkoreit, J., Jones, L., Gomez, A.N.,
  Kaiser, L.u., Polosukhin, I.: Attention is all you need. In: Guyon, I.,
  Luxburg, U.V., Bengio, S., Wallach, H., Fergus, R., Vishwanathan, S.,
  Garnett, R. (eds.) Advances in Neural Information Processing Systems 30, pp.
  5998--6008. Curran Associates, Inc. (2017),
  \url{http://papers.nips.cc/paper/7181-attention-is-all-you-need.pdf}

\bibitem{Voorhees94}
Voorhees, E.M.: Query expansion using lexical-semantic relations. In: Proc. of
  the 17th Annual International ACM SIGIR Conference on Research and
  Development in Information Retrieval. pp. 61--69. SIGIR '94, Springer-Verlag
  New York, Inc., New York, NY, USA (1994),
  \url{http://dl.acm.org/citation.cfm?id=188490.188508}

\bibitem{NeuralHype2}
Yang, W., Lu, K., Yang, P., Lin, J.: Critically examining the "neural hype":
  Weak baselines and the additivity of effectiveness gains from neural ranking
  models. In: Piwowarski, B., Chevalier, M., Gaussier, {\'{E}}., Maarek, Y.,
  Nie, J., Scholer, F. (eds.) Proceedings of the 42nd International {ACM}
  {SIGIR} Conference on Research and Development in Information Retrieval,
  {SIGIR} 2019, Paris, France, July 21-25, 2019. pp. 1129--1132. {ACM} (2019).
  \doi{10.1145/3331184.3331340}, \url{https://doi.org/10.1145/3331184.3331340}

\bibitem{Zhai2001}
Zhai, C., Lafferty, J.D.: A study of smoothing methods for language models
  applied to ad hoc information retrieval. In: Proc. of the SIGIR conference.
  pp. 334--342 (2001)

\end{thebibliography}

\end{document}